# UML 2.0 - Overview and Perspectives in SoC Design

Tim Schattkowsky
University of Paderborn/C-LAB
Paderborn, Germany

**Abstract**

*The design productivity gap requires more efficient design methods. Software systems have faced the same challenge and seem to have mastered it with the introduction of more abstract design methods. The UML has become the standard for software systems modeling and thus the foundation of new design methods. Although the UML is defined as a general purpose modeling language, its application to hardware and hardware/software codesign is very limited. In order to successfully apply the UML at these fields, it is essential to understand its capabilities and to map it to a new domain.*

## 1. Introduction

Today, electronic systems design has to cope with shortened product cycles and steadily increasing system complexity that are not yet complemented by a comparable increase in design productivity. Thus, the design productivity gap has become one of the most important problems in electronic systems design. Design methods allowing for increased abstraction as well as better reuse and integration of IPs are necessary to cope with this challenge.

The Unified Modeling Language (UML) has been successfully applied in software systems engineering as well as in some other domains. It has the capabilities to be applied to the design of hardware systems as well. This also promises an easy path to hardware/software codesign through integration with existing practice in software design. Furthermore, the notation has to be integrated with a design process to get a complete design method like in MDA. However, it is first necessary to understand, which capabilities the UML provides and how these can be employed in EDA design flows.

## 2. The Unified Modeling Language (UML)

The UML is a general purpose modeling language that marks the result of the unification of elements from three of the most significant object oriented design approaches in the later 90s. UML 1.x essentially combined elements from Booch [1], OOSE [3] and OMT [9]. Furthermore, it was influenced by the ROOM method that later evolved into UML-RT [10], a real-time profile for UML. At present, the UML 2.0 specification is on its way to finalization by the OMG. UML 2.0 includes significant enhancements in all areas. It covers 13 diagram types to describe various structural, behavioral and physical aspects of a system.

The different diagrams serve different purposes during the development process. It is important to note that the UML only provides a multi purpose notation that needs to be accompanied by a development process. This process defines the actual application variants of a particular UML diagram type.

The *Class Diagram* is probably the most well known diagram. It describes structural aspects of a system in terms of classes and associations between those classes. A class may have attributes and operations. Furthermore, interfaces and generalizations can be used to create object oriented type hierarchies. Instances of a Class Diagram are called an *Object Diagram* and describe how individual class instances (objects) are related.

*Package Diagrams* can be used to define packages, which provide just a little more than a namespace for classes that can be composed from other packages and new classes.

Behavioral specification in the UML at the highest level often starts by the identification of the use cases for a system described in terms of involved actors. This is depicted in *Use Case Diagrams*. However, detailed behavioral specifications usually rely on *State Machine Diagrams* that represent the UML StateChart [2] variant and Activity Diagrams for describing control oriented behavior (e.g., in states of a State Machine Diagram). UML 2.0 introduces token semantics for these Activity Diagrams that move them semantically close to high-level Petri Nets. Furthermore, *Interaction diagrams* can be used to model interactions as traces of exchanged messages. These diagrams come in no less than four flavors including the *Sequence Diagram*, which has been extended in UML 2.0 to be comparable to an SDL Message Sequence Chart (MSC).

Finally, the *Component Diagram* and the *Deployment Diagram* can be used to describe the composition and physical deployment of a system.

Together, those diagrams can be used to provide a detailed model of a complete system at various levels of abstraction (e.g., starting from use cases down to activities describing every detail of a certain behavior).

However, the UML just provides a general notation that fits many purposes. It must be tailored to be effectively applied to a certain domain. The relevant model elements of the UML need to be identified and may be refined into a set of domain specific subtypes. This is achieved using a UML *profile* that defines a relevant domain-specific UML subset with semantic extensions for the supported model elements. On example for such profiles is the aforementioned UML-RT profile.

Thus, to apply UML to SoC design, it is important to define such a domain specific subset of the UML and its semantics as well as the diagram types to be used.

## 3. Model Driven Architecture (MDA)

The OMG MDA is essentially based on providing system specifications as a Platform Independent Model (PIM), which is to be more or less automatically transformed to a Platform Specific Model (PSM) for a different platform using a platform-specific mapping. This can be regarded as an enhancement to platform-based design. Both models are UML models at different levels of abstraction. The Platform Specific Model is then used for complete code generation. However, how these steps are accomplished is up to the actual MDA tool.

There already exist methods that follow the MDA approach. The Executable UML (xUML) approach [8] includes a complete development methodology. It is based on the Action Specification Language (ASL), which is essential as it describes notation and semantics for single actions like operation calls and assignments in UML models and thus closes the last gap to complete system specification. ASL has been integrated into the OMG-adopted UML Standard for precise action semantics [5] and is also used in the eXecutable and Translatable UML approach ($^X_T$UML) which has been developed by Project Technology based on Mellor's approach to executable UML [4]. Both approaches use code generators to produce the actual implementations from the UML models. However, the application of such code generation for hardware descriptions still needs to be demonstrated.

## 4. Perspectives for SoC Design

The UML is quite appealing for application in hardware design, as digital hardware and software have quite a lot in common. At a reasonable level of abstraction, both already have similar structural (e.g., software components and IP cores) and behavioral models (e.g., StateCharts, Petri Nets). Application of UML to hardware systems promises to close the design gap between hardware and software systems by providing inherent interfaces that should lead to full interchangeability between hardware and software. Furthermore, the early prototyping and inherent software simulation capabilities of such an approach are appealing, as they promise cost and time savings. Finally, the application of the MDA concept also to hardware finally promises large scale reuse and portability.

However, while the UML is by definition a general purpose modeling language, it cannot be instantly applied to hardware design. As it is just a notation, meaning must be given to all the relevant language elements. It is necessary to the tailor the UML in a way that the domain specific significant requirements like seamless integration of existing IP can be met.

Thus, the real world things that need to be represented have to be identified and consistently put into the right context as UML model elements. The relation between the concepts used in the UML and real circuits has to be clarified. For example, the notion of class, object and component have to be aligned in this context and the structural semantics of operations need to be defined.

Furthermore, the application of the different diagrams in the design process needs to be clarified. Many ideas can be derived from the existing methods for software and hardware design. In the context of MDA, the integration with a design process can be achieved.